\documentclass[pre,eqsecnum,twocolumn, showpacs,nofootinbib]{revtex4}
\global\arraycolsep=2pt
\usepackage{stmaryrd}
\usepackage{amsmath}
\usepackage{amssymb}
\usepackage{graphicx}
\usepackage{textcomp}
\usepackage{calrsfs}
\usepackage{yfonts}
\usepackage{bm}

\newcommand{\Tr}{\mathop{\mbox{Tr}}}

\newcommand{\be}{\begin{equation}}                                                
\newcommand{\ee}{\end{equation}}
\newcommand{\ben}{\begin{equation*}}                                              
\newcommand{\een}{\end{equation*}}
\newcommand{\bea}{\begin{eqnarray}}                                               
\newcommand{\eea}{\end{eqnarray}}

\newcommand{\nn}{\nonumber}
\begin{document}

\title{Investigations of the torque anomaly in an annular sector. II. 
Global calculations, electromagnetic case}

\date{\today}

\author{Kimball A. Milton}\email{milton@nhn.ou.edu}
\author{Prachi Parashar}\email{prachi@nhn.ou.edu}
\author{E. K.  Abalo}\email{abalo@nhn.ou.edu}
\affiliation{Homer L. Dodge  Department of
Physics and Astronomy, The University of Oklahoma, Norman, OK 73019-2053, USA}
 \author{Fardin Kheirandish}\email{fkheirandish@yahoo.com}
\affiliation{ 
Department of Physics, Faculty of Science, University of Isfahan,
Hezar-Jarib St., 81746-73441,
Isfahan, Iran}
 \author{Klaus Kirsten} \email{Klaus_Kirsten@baylor.edu}
 \affiliation{Department of Mathematics, Baylor University, One 
Bear Place, Waco, TX 76798-7328, USA}

\begin{abstract}
Recently, it was suggested that there was some sort of
breakdown of quantum field theory in the presence of boundaries,
manifesting itself as a torque anomaly.  In particular, Fulling et al.\
used the finite energy-momentum-stress tensor in the presence
of a perfectly conducting wedge, calculated many years ago by Deutsch and
Candelas, to compute the torque on one of the wedge boundaries,
where the latter was cutoff by integrating the torque density down to minimum
lower radius greater than zero. They observed that that torque
 is not equal to the negative derivative
of the energy obtained by integrating the 
energy density down to the  same minimum
radius.  This motivated a calculation of the torque and energy
in an annular sector obtained by the intersection of the wedge with two
coaxial cylinders.  In a previous paper we showed that for the analogous
scalar case, which also exhibited a torque anomaly in the absence of the
cylindrical boundaries, the point-split regulated torque and energy indeed
exhibit an anomaly, unless the point-splitting is along the axis direction.
In any case, because of curvature divergences, no unambiguous finite part can
be extracted.  However, that ambiguity is linear in the wedge angle; if
the condition is imposed that the linear term be removed, the resulting
torque and energy is finite, and exhibits no anomaly.  In this paper,
we demonstrate the same phenomenon takes place for the electromagnetic
field, so there is no torque anomaly present here either.  This is a
nontrivial generalization, since the anomaly found by Fulling et al.\ is linear
for the Dirichlet scalar case, but nonlinear for the conducting electromagnetic
case.
\end{abstract}
\maketitle
\section{Introduction}
Recently, Fulling et al.~\cite{fulling12} 
suggested that a quantum torque anomaly
exists in field theories in the presence of boundaries.  This is related,
but somewhat distinct from that group's 
earlier discussion of a pressure anomaly
\cite{Estrada:2012yn}, since the latter explicitly depended on taking
seriously the distance dependence of stress tensor components below the cutoff
scale.  In the new torque anomaly, the stress tensor employed is the completely
finite one (cutoff independent) for an ideal wedge 
calculated first by Dowker and Kennedy \cite{Dowker:1978md}
for the Dirichlet scalar case, and then given for electromagnetic fields 
subject to perfectly conducting boundaries by Deutsch and Candelas
\cite{deutsch}.  These computations were later revisited by Brevik and Lygren
\cite{brevik} and by Saharian and Tarloyan
\cite{saharian}.  It should, however, be borne
in mind that in computing those completely finite vacuum expectation values
of the stress tensor, regularization, such as afforded by point-splitting
in the angular or the radial direction, is necessary, before the subtraction
of the free-space vacuum stress tensor is effected. So the distinction 
between the two types of anomalies is not so sharp.

Naturally, the stress tensor computed for the wedge is singular at the apex
of the wedge.  Therefore, it is not possible to compute the total energy of
the wedge, or the torque exerted by quantum fluctuations of the interior
fields on one of the sides of the wedge.  So what is proposed in 
Ref.~\cite{fulling12} is to integrate only from some nonzero inner radius $a$
from the apex, for both the torque and the energy.  That is, let the torque
per unit length be
\be
\tau(a,\alpha)=\int_a^\infty d\rho\,\rho\,\langle T^{\theta}{}_\theta\rangle,
\ee
where the integral is over one of the wedge sides, $\theta$ is the axial
angle, and $\alpha$ is the angle of the wedge. 
The corresponding energy per unit length is
\be
\mathcal{E}(a,\alpha)=\int_a^\infty d\rho\,\rho\, \int_0^\alpha d\theta 
\,\langle T^{00}\rangle,
\ee
It is immediately seen from
the Deutsch-Candelas stress tensor that
\be
\tau(a,\alpha)\ne-\frac\partial{\partial\alpha}\mathcal{E}(a,\alpha).
\ee
This is Fulling's torque anomaly.

\begin{figure}
 \begin{center}
  \includegraphics{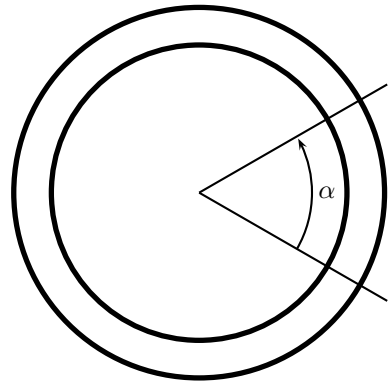}
\caption{\label{fig:annpist} The Casimir energy and torque are 
calculated for the region bounded between two perfectly conducting 
cylinders, of radius $a$ and $b$, bounded in the angular direction 
by two perfectly conducting radial planes, making an angle $\alpha$
between them.}
 \end{center}
\end{figure}

A possible resolution of this anomaly has been suggested by Dowker
\cite{dowker13}.  It would appear that what is necessary is more than
simply putting in spatial cutoffs on the integrals. This, in effect,
equates the force on a semi-infinite plate, not touching a second semi-infinite
plate,  with the negative derivative of the quantum vacuum energy contained
in only the open region between those plates, rather than the energy in all
of space.  Therefore, we here are considering a region completely bounded by
conducting surfaces:  the two radial wedge boundaries and two circular
cylindrical boundaries sharing a common axis, as shown in 
Fig.~\ref{fig:annpist}.  In Ref.~\cite{milton13} we considered such a
geometry for a massless scalar field, with Dirichlet boundaries.
We regulate the integral by point-splitting in the time or the axial direction.
For the former, the divergent expressions indeed exhibit an anomaly, in
that the torque is not equal to the negative derivative of the energy
contained within the sector.  This anomaly disappears for point-splitting
in the axial direction, consistent with the findings of 
Ref.~\cite{Estrada:2012yn}, since that is a neutral direction, not referring
to the stress tensor  components involved in either the energy density or 
the torque density.  Introducing the cylindrical boundaries, however, 
causes another problem by generating divergences 
associated with curvature.  These
curvature divergences generate logarithmic terms in the cutoff parameter,
which means that it is impossible to extract a finite energy.  However, all
the divergences encountered are linear functions of the wedge angle,
so if we demand that the ``renormalized'' observable energy approach zero
as the wedge angle gets large, we can remove such terms, yielding a finite
energy which indeed has the correct balance with the torque.
These results are consistent with the annular piston results calculated
a few years ago \cite{Milton:2009bz}, using the multiple-scattering technique.

In the present paper, we generalize the result of Ref.~\cite{milton13},
hereafter referred to as I,  to
the electromagnetic situation, with perfect conducting boundary conditions.
In the next section we set up the general Green's dyadic formulation,
for the situation of cylindrical symmetry, where, with perfectly conducting
boundaries, we have the complete decomposition between TE and TM modes.
This means that the TM modes are the Dirichlet modes calculated in I, 
while the TE modes are scalar Neumann modes.  In 
Sec.~\ref{sec:annsec} we derive formulas for the energy in the sector,
as well as the torque on one of the radial planes.  These quantities are 
regulated by point splitting either in the temporal or the axial direction.
All the divergent
terms are extracted for the energy in Sec.~\ref{sec:div}, proportional to
the volume, the surface area, the corners, and curvature corrections.  These
correspond to known terms in the heat kernel expansion for this problem
\cite{Dowker:1995sp,apps,Nesterenko:2002ng}.  The finite part is extracted in 
Sec.~\ref{sec:finite}, which arises from the uniform asymptotic expansion
of the Bessel functions appearing in the Green's functions, and the remainder,
which is computed numerically in Sec.~\ref{sec:num}.  Just as in the scalar
case, the numerical results exhibit a linear dependence on the wedge angle
for sufficiently (not very) large angles.  So it is proposed to remove
this linear dependence completely, by a renormalization process that
eliminates all the divergent terms,
leaving finite results which satisfy the expected balance between energy and
torque.  Concluding remarks are offered in Sec.~\ref{sec:concl}.

\section{Green's dyadic}
\label{sec:gd}

The electromagnetic Feynman Green's  dyadic, which corresponds to 
the vacuum expectation value of the time-ordered product of electric
fields, satisfies the differential equation
\be
\left(\frac1{\omega^2}\bm{\nabla}\times\bm{\nabla} \times-\bm{1}\right)\bm{\Gamma}
(\mathbf{r,r'};\omega)=\bm{1}\delta(\mathbf{r-r'}),
\ee
or, for the divergenceless dyadic $\bm{\Gamma}'=\bm{\Gamma}+\bm{1}$,
\be
\left(\frac1{\omega^2}\bm{\nabla}\times\bm{\nabla} 
\times-\bm{1}\right)\bm{\Gamma}'
(\mathbf{r,r'};\omega)=\frac1{\omega^2}\bm{\nabla}
\times(\bm{\nabla}\times\bm{1})\delta(\mathbf{r-r'}).
\ee
Here, and in the following, we have taken a Fourier transform in time.
Henceforth, we will suppress the explicit reference to the frequency
dependence.
For a situation with cylindrical symmetry, and perfect conducting boundary
conditions, the modes decouple into transverse electric and 
transverse magnetic modes, and we can write
\be
\bm{\Gamma}'=\bm{E}G^E+\bm{H}G^H,\label{gammaeh}
\ee
in terms of transverse electric and magnetic Green's functions, where the
polarization tensors have the structure (for example, see Ref.~\cite{brevik})
\begin{subequations}
\bea
\bm{E}&=&-\nabla^2(\bm{\nabla}\times\bm{\hat z})
(\bm{\nabla}'\times\bm{\hat z}),\\
\bm{H}&=&(\bm{\nabla}\times(\bm{\nabla}\times\bm{\hat z}))
(\bm{\nabla}'\times(\bm{\nabla}'\times\bm{\hat z})),
\eea
\end{subequations}
where $z$ is the translationally invariant direction.
Acting on a completely translationally invariant function,
\be
\bm{E+H}=-\nabla_\perp^2(\bm{\nabla\nabla}-\bm{1}\nabla^2),
\ee
where
\be
\nabla^2=\nabla_\perp^2+\frac{\partial^2}{\partial z^2}.
\ee
Further useful properties of $\bm{E}$ and $\bm{H}$ are
\begin{subequations}
\be
\bm{\nabla}\times\bm{E}\,\times\loarrow{\bm{\nabla}}'=
\bm{H}\nabla^{2},\quad
\bm{\nabla}\times\bm{H}\,\times\loarrow{\bm{\nabla}}'=
\bm{E}\nabla^{\prime 2},
\ee
where it is understood that both gradients act on everything to the
right, and
\bea
\bm{E}(\mathbf{r,r'})\cdot\bm{H}(\mathbf{r'',r'''})
&=&\bm{H}(\mathbf{r,r'})\cdot\bm{E}(\mathbf{r'',r'''})=0,\nn\\
\\
\bm{E}(\mathbf{r,r'})\cdot\bm{E}(\mathbf{r'',r'''})
&=&\bm{E}(\mathbf{r,r'''})\nabla^{\prime2}_\perp
\nabla^{\prime\prime2},\\
\bm{H}(\mathbf{r,r'})\cdot\bm{H}(\mathbf{r'',r'''})&=&\bm{H}(\mathbf{r,r'''})
\nabla^{\prime2}_\perp\nabla^{\prime\prime2},
\eea
\end{subequations}
where we will understand that after differentiation, the intermediate 
coordinates $\mathbf{r'}$ and $\mathbf{r''}$ become identified.

For electromagnetism, the energy density is
\be
u=T^{00}=\frac{E^2+B^2}2,
\ee
so by use of the Maxwell equations the energy contained in a volume $V$
with perfectly conducting boundaries $\partial V$
 becomes, in terms of the
imaginary frequency $\zeta=-i\omega$,
\bea
\int_V(d\mathbf{r}) u(\mathbf{r})
&=&
\frac12\int_V(d\mathbf{r})\Tr \left[\bm{1}+\frac1{\zeta^2}\left(\nabla^2\bm{1}-
\bm{\nabla\nabla}\right)\right]\nn\\
&&\quad\cdot \mathbf{E(r) E(r')^*}
\bigg|_{\mathbf{r'=r}},
\eea
because
\bea
&&\int_V(d\mathbf{r})\Tr\bm{\nabla}\times\bm{\nabla}\times [\mathbf{E}
(\mathbf{r})\mathbf{E}(\mathbf{r})^*]\nn\\
&=&i\omega \oint_{\partial V}\sigma\mathbf{
\hat n\times B(\mathbf{r})\cdot E(\mathbf{r})^*}=0,
\eea
provided the boundaries are perfect conductors.
Quantum mechanically, we replace the expectation value of the 
product of electric fields $\mathbf{E(r)}$ by the Green's dyadic:
\be
\langle \mathbf{E}(\mathbf{r})\mathbf{E}(\mathbf{r}')^*\rangle=\frac1i
\bm{\Gamma}(\mathbf{r,r'}).
\ee
Because we will be regulating all integrals by point splitting,
we can ignore delta functions (contact terms) in evaluations,
so in terms of $\bm{\Gamma}'$, the quantum vacuum energy is
\bea
E&=&\int_V(d\mathbf{r})\int_{-\infty}^\infty \frac{d\omega}{2\pi}
e^{-i\omega t}
 \langle u(\mathbf{r})\rangle\nn\\
&=&\frac1{2i}\int_V(d\mathbf{r})\Tr \int_{-\infty}^\infty \frac{d\omega}{2\pi}
e^{-i\omega t}\frac1{\zeta^2}(\nabla^2+\zeta^2)\bm{\Gamma}'(\mathbf{r,r'})
\big|_{\mathbf{r'}\to\mathbf{r}}\nn\\
&=&\int_V(d\mathbf{r})\int\frac{d\zeta}{2\pi}e^{i\zeta t_E}
\Tr\bm{\Gamma}'(\mathbf{r,r}),
\eea
where in the last equation we have performed the rotation to Euclidean space,
so $-it\to t_E$ is a Euclidean time-splitting parameter, going to zero through 
positive
values.  This is a well-known formula, for example, see 
Ref.~\cite{Milton:2010yw}.  The energy may be written 
in terms of the scalar Green's functions in Eq.~(\ref{gammaeh}),
\be
E=\int_V(d\mathbf{r})\int\frac{d\zeta}{2\pi}e^{i\zeta t_E}\zeta^2\nabla_\perp^2(G^E+G^H)
(\mathbf{r,r'})\big|_{\mathbf{r'\to r}},
\label{egegh}
\ee
which again involves an integration by parts, and use of the
perfect conducting boundary conditions on both arguments of the Green's
functions (see below)
\be
\oint_{\partial V}d\bm{\sigma}\cdot\bm{\nabla}_\perp G^{E,H}_\perp(
\mathbf{r,r'})\bigg|_{\mathbf{r'\to r}}=0.
\ee
The decomposition theorems contained in this section are familiar from
waveguide theory, for example, see Ref.~\cite{Schwinger}.
\section{Annular Sector}\label{sec:annsec}
We now specialize to the situation at hand, an annular sector bounded by two
concentric cylinders, intercut by a co-axial wedge, as 
illustrated in Fig.~\ref{fig:annpist}.
The inner cylinder has radius $a$, the outer $b$, and the wedge angle is 
$\alpha$.  
The axial direction is chosen to coincide with the $z$ axis.
The explicit form for the Green's dyadic is
\bea
\bm{\Gamma}'(\mathbf{r,r'})&=&-\frac2\alpha\sum_m\int_{-\infty}^\infty 
\frac{dk}{2\pi} e^{ik(z-z')}\frac1{\kappa^2}\nn\\
&&\times\bigg[\bm{E}(\mathbf{r,r'})\cos \nu\theta\cos \nu\theta' g_\nu^E
(\rho,\rho')\nn\\
&&\mbox{}+\bm{H}(\mathbf{r,r'})\sin \nu\theta\sin \nu\theta' g_\nu^H
(\rho,\rho')\bigg].
\eea
Here $\nu=mp$ where $p=\pi/\alpha$, and $\kappa^2=\zeta^2+k^2$.  
The $m$ summation runs from 0 to $\infty$ for the TE modes, but
only from 1 to $\infty$ for the TM modes.  We
will  see the crucial role of the TE ``zero mode'' in the following.
The H mode vanishes on the radial planes,
and on the circular arcs,
\be
g_\nu^H(a,\rho')=g^H_\nu(b,\rho')=0.\label{dbc}
\ee
The normal derivative of the E mode vanishes on the radial planes, as
it does on the circular arcs:
\be
\frac{\partial}{\partial\rho}g^E_\nu(\rho,\rho')\bigg|_{\rho=a,b}=0.\label{nbc}
\ee
Thus, the TE mode corresponds to a scalar mode satisfying Neumann boundary
conditions, while the TM modes correspond to  scalar Dirichlet modes.
Therefore, the latter are exactly those found in the corresponding scalar
calculation in I.  Both scalar Green's functions satisfy the same equation:
\be
\left(-\frac1\rho\frac\partial{\partial\rho}\rho\frac\partial{\partial\rho}
+\kappa^2+\frac{\nu^2}{\rho^2}\right)g^{E,H}_\nu=\frac1\rho\delta(\rho-\rho').
\ee
Therefore, imposing the boundary conditions (\ref{dbc}) and (\ref{nbc}) we find
\begin{subequations}
\bea
g^H_\nu(\rho,\rho')&=&I_\nu(\kappa\rho_<)K_\nu(\kappa\rho_>)\nn\\
&&\mbox{}-\frac{K_\nu(\kappa a)K_\nu(\kappa b)}{\Delta}I_\nu(\kappa\rho)
I_\nu(\kappa\rho')\nn\\
&&\mbox{}-\frac{I_\nu(\kappa a)I_\nu(\kappa b)}{\Delta}
K_\nu(\kappa\rho)K_\nu(\kappa\rho')\nn\\
&&\mbox{}+\frac{I_\nu(\kappa a)K_\nu(\kappa b)}{\Delta}
[I_\nu(\kappa\rho)K_\nu(\kappa\rho')\nn\\
&&\quad\mbox{}+K_\nu(\kappa\rho)
I_\nu(\kappa\rho')],\\
g^E_\nu(\rho,\rho')&=&I_\nu(\kappa\rho_<)K_\nu(\kappa\rho_>)\nn\\
&&\mbox{}-\frac{K'_\nu(\kappa a)K'_\nu(\kappa b)}{\hat\Delta}
I_\nu(\kappa\rho)
I_\nu(\kappa\rho')\nn\\
&&\mbox{}-\frac{I'_\nu(\kappa a)I'_\nu(\kappa b)}{\hat\Delta}
K_\nu(\kappa\rho)K_\nu(\kappa\rho')\nn\\
&&\mbox{}+\frac{I'_\nu(\kappa a)K'_\nu(\kappa b)}{\hat\Delta}
[I_\nu(\kappa\rho)K_\nu(\kappa\rho')\nn\\
&&\quad\mbox{}+K_\nu(\kappa\rho)
I_\nu(\kappa\rho')],
\eea
\end{subequations}
where
\begin{subequations}\label{Deltas}
\bea
\Delta_\nu(\kappa a,\kappa b)&=&I_\nu(\kappa b)K_\nu(\kappa a)-I_\nu(\kappa a)
K_\nu(\kappa b),\\
\hat\Delta_\nu(\kappa a,\kappa b)&=&I'_\nu(\kappa b)K'_\nu(\kappa a)
-I'_\nu(\kappa a)K'_\nu(\kappa b).\label{deltat}
\eea
\end{subequations}

\subsection{Energy}

Now using Eq.~(\ref{egegh}) we have for the energy per length in the $z$
direction
\bea
\mathcal{E}&=&-\int\frac{d\zeta}{2\pi}\frac{dk}{2\pi}\zeta^2 e^{i\zeta t_E}
e^{ikZ}\nn\\
&&\quad\times\sum_m\!\!\int_a^bd\rho\,\rho[g^E_\nu(\rho,\rho)+g^H_\nu(\rho,\rho)].
\eea
In I we showed that 
\be
\int_a^b d\rho\,\rho\,g^H_\nu(\rho,\rho) = \frac1{2\kappa}
\frac\partial{\partial\kappa}\ln\Delta,\ee
and in just the same way we can show \cite{prudnikov}
\be
\int_a^b d\rho\,\rho\,g^E_\nu(\rho,\rho)=\frac1{2\kappa}\frac\partial{\partial
\kappa}\ln\kappa^2\hat\Delta,
\ee in terms of the quantities defined in Eq.~(\ref{Deltas}).  Therefore,
the energy per unit length is given by
\be
\mathcal{E}=-\frac1{4\pi}\int_0^\infty d\kappa\,\kappa^2 f(\kappa\delta,\phi)
\sum_m\frac\partial{\partial\kappa}\ln\kappa^2\Delta\hat\Delta.\label{emen}
\ee
Here, to explore the effects of different point-splitting schemes,
we write
\be
\zeta=\kappa\cos\gamma,\quad k=\kappa\sin\gamma,\quad t_E=\delta\cos\phi,
\quad Z=\delta\sin\phi,
\ee
where $Z=z-z'$ is an infinitesimal point splitting in the $z$ direction,
and then
 we define the regulator function
\be
f(\kappa\delta,\phi)=\int_0^{2\pi}\frac{d\gamma}{2\pi}\cos^2\gamma 
\,e^{i\kappa\delta\cos(\gamma-\phi)},
\ee
which equals 1/2 for $\delta=0$.  For finite $\delta$, temporal splitting
corresponds to
\begin{subequations}
\be
f(\kappa\delta,0)=J_0(\kappa\delta)-\frac1{\kappa\delta}J_1(\kappa\delta),
\ee
while $z$-splitting corresponds to
\be
f(\kappa\delta,\pi/2)=\frac1{\kappa\delta}J_1(\kappa\delta).
\ee
\end{subequations}

\subsection{Torque}\label{sec:torque}
To compute the torque on one of the radial planes, we need to compute
the angular component of the stress tensor,
\bea
\langle T^{\theta}{}_{\theta}\rangle
&=&-\frac12\langle E_\theta^2-B_\rho^2-B_z^2
\rangle\nn\\
&=&-\frac1{2i}\bigg[\bm{\hat\theta}\cdot \bm{\Gamma}'\cdot\bm{\hat\theta}
+\frac1{\omega^2}\bm{\hat\rho\cdot\nabla\times\Gamma'\,
\times\loarrow
\nabla'\cdot\hat\rho}\nn\\
&&\quad+\frac1{\omega^2}\bm{\hat z\cdot\nabla\times\Gamma'\,\times
\loarrow\nabla'\cdot \hat z}\bigg]\bigg|_{\mathbf{r'\to r}}.
\eea
The torque then is immediately obtained by integrating the first moment of 
this over one radial side of the annular region, that is, for $\theta=0$ or
$\alpha$:
\bea
\tau&=&\int_a^b d\rho\,\rho\,\int_{-\infty}^\infty \frac{d\omega}{2\pi} e^{-i
\omega(t-t')} \langle T^{\theta}{}_\theta\rangle\nn\\
&=&\frac1\alpha\sum_m\nu^2\!\!\int_0^\infty \frac{d\kappa\,\kappa}{2\pi}
J_0(\kappa\delta)\!\!\int_a^b\frac{d\rho}\rho[g_\nu^E(\rho,\rho)
+g_\nu^H(\rho,\rho)].\nn\\
\eea
In I we gave the radial integral for the TM part:
\be
\int_a^b \frac{d\rho}{\rho}g_\nu^H(\rho,\rho)=-\frac\alpha{2\nu^2}\frac
\partial{\partial\alpha}\ln\Delta,
\ee
and we can show the same relation holds for the TE part \cite{prudnikov}:
\be
\int_a^b \frac{d\rho}{\rho}g_\nu^E(\rho,\rho)=-\frac\alpha{2\nu^2}\frac
\partial{\partial\alpha}\ln\hat\Delta.
\ee
Thus the electromagnetic torque on one of the planes is
\be
\tau=-\frac\partial{\partial\alpha}\frac1{4\pi}\sum_m\int_0^\infty
d\kappa\kappa\,J_0(\kappa\delta)\ln\kappa^2\Delta\hat\Delta.
\ee
Using integration by parts in Eq.~(\ref{emen}), and Bessel's equation, we see 
this is indeed the negative derivative with respect to the wedge angle of the
interior energy provided $\phi=\pi/2$, that is, for point-splitting in the
$z$ direction.  We will now proceed to evaluate the energy, by explicitly
isolating the divergent contributions as $\delta\to 0$, and extract the
finite parts.  Will it be true, as in the scalar case, that after 
renormalization the finite torque is equal to the negative derivative of
the finite energy with respect to the wedge angle?

\section{Divergent terms for the TE energy}\label{sec:div}
We now turn to the examination of the Neumann or TE contribution to the
Casimir energy of the annular region, which is 
\be
\hat{\mathcal{E}}=-\frac1{4\pi}\int_0^\infty d\kappa\,\kappa^2 
f(\kappa\delta,\phi)
\sum_{m=0}^\infty \frac\partial{\partial \kappa}\ln\kappa^2\hat\Delta,
\label{energyt}
\ee
where $\hat\Delta$ is given by Eq.~(\ref{deltat}).
As in the Dirichlet case, we expand the Bessel functions according to the
uniform asymptotic expansion, which here reads
\begin{subequations}
\label{uae}
\bea
I_\nu'(\nu \xi)&\sim&\frac1{\sqrt{2\pi\nu t}}\frac1\xi e^{\eta\nu}\left(1+
\sum_{k=1}^\infty\frac{v_k(t)}{\nu^k}\right),\\
K_\nu'(\nu \xi)&\sim&-\sqrt{\frac{\pi}{2\nu t}}\frac1\xi e^{-\eta\nu}\left(1+
\sum_{k=1}^\infty(-1)^k\frac{v_k(t)}{\nu^k}\right),\nn\\
\eea
\end{subequations}
where\footnote{The variable $\xi$ is the same as that called $z$ in I; we
have changed the notation here to avoid confusions with the axial coordinate.}
 $t=(1+\xi^2)^{-1/2}$, $d\eta/d\xi=1/(\xi t)$, 
and the polynomials $v_k(t)$ 
are generated from those for the functions $I_\nu$ and $K_\nu$ by
\be
v_0(t)=1,\,\, v_k(t)=u_k(t)+t(t^2-1)\left[\frac12 u_{k-1}(t)+t u_{k-1}'(t)
\right].
\ee
Because of this behavior, the second product of Bessel functions in 
Eq.~(\ref{deltat}) is exponentially subdominant.  Thus the logarithm in
Eq.~(\ref{energyt}) is asymptotically
\bea
\ln\kappa^2\hat\Delta&\sim&\mbox{constant}+\nu[\eta(\xi)-\eta(\tilde \xi)]
+\left(t^{-1/2}+\tilde t^{-1/2}
\right)\nn\\
&&\mbox{}+\ln\left(1+\sum_{k=1}^\infty \frac{v_k(t)}{\nu^k}\right)\nn\\
&&\mbox{}+\ln\left(1+\sum_{k=1}^\infty (-1)^k\frac{v_k(\tilde t)}{\nu^k}
\right),
\eea
where $\xi=\kappa b/\nu$, $\tilde \xi=\xi a/b$,  
$\tilde t=(1+\tilde \xi^2)^{-1/2}$.
Here the constant means a term independent of $\kappa$, which will not
survive differentiation.  Note that the $1/\xi$ behavior seen in the
prefactors in Eq.~(\ref{uae}) 
is canceled by the multiplication of $\hat\Delta$
by $\kappa^2$.  In the following, we will consider the $z$-splitting regulator,
$\phi=\pi/2$, since the result for time-splitting may be obtained
by differentiation:
\be
\hat{\mathcal E}(0)=\frac{\partial}{\partial\delta}[\delta\hat{\mathcal{E}
}(\pi/2)].\label{relttoz}
\ee

We now extract the divergences, that is the terms proportional  to nonpositive
powers of $\delta$, just as in I.  We label those terms by the corresponding
power of $1/\delta$. The calculation closely parallels that in I, except for
the additional zero mode, $m=0$.  Except for that term, the leading divergence
is exactly that found in I, 
\be
\hat{\mathcal{E}}_4^{m>0}=-\frac{\alpha(b^2-a^2)}{4\pi^2\delta^4}+\frac{b-a}
{8\pi\delta^3}.\label{et4}
\ee
However, the $m=0$ term yields 
\be
\hat{\mathcal{E}}_4^{m=0}=-\frac{b-a}{4\pi\delta^3},
\ee
thereby (correctly) reversing the sign of the second term in Eq.~(\ref{et4}).
Thus the leading divergence is again the expected Weyl volume divergence:
\be
\hat{\mathcal{E}}^{(4)}=-\frac{A}{2\pi^2\delta^4},\quad A=\frac12\alpha
(b^2-a^2).\ee

Evidently, the $O(\nu^{-3})$ term, for $m>0$, is exactly reversed in sign
from that for the Dirichlet term,
\be
\hat{\mathcal{E}}_3^{m>0}=-\frac{\alpha(a+b)}{16\pi\delta^3}
+\frac1{8\pi\delta^2},\ee
but again the sign of the subleading term is reversed by including $m=0$:
\be
\hat{\mathcal{E}}_3^{m=0}=-\frac1{4\pi\delta^2}.
\ee
Thus, we get the correct surface area and corner terms:
\begin{subequations}
\bea
\hat{\mathcal{E}}^{(3)}&=&-\frac{P}{16\pi\delta^3}, \quad P=\alpha(a+b)+
2(b-a),\\
\hat{\mathcal{E}}^{(2)}&=&-\frac{C}{48\pi\delta^2},\quad C=
4\left(\frac\pi{\pi/2}-\frac{\pi/2}\pi\right)=6.\nn\\
\eea
\end{subequations}

Closely following the path blazed in computing the divergent terms coming
from the polynomial asymptotic corrections in the Dirichlet case in I, but
including the $m=0$ terms, we find the first three curvature corrections
\begin{subequations}
\bea
\hat{\mathcal{E}}_2&=&\frac3{64\pi}\frac1\delta\left(\frac1a-\frac1b\right),
\\
\hat{\mathcal{E}}_1&=&-\frac{5}{1024}\frac\alpha\pi
\frac1\delta\left(\frac1a+\frac1b
\right)+\frac{3\ln\delta}{128\pi}\left(\frac1{a^2}+\frac1{b^2}\right).\nn\\
\eea
\end{subequations}

\subsection{$m=0$ case}
Before proceeding, it is time to recognize that use of the uniform asymptotic
expansion is apparently
 inconsistent for $m=0$, because $\nu=0$ then.  So let us
calculate the $m=0$ contribution directly from
\bea
\hat{\mathcal{E}}_{m=0}&=&-\frac1{4\pi}\int_0^\infty d\kappa\,\kappa^2
\frac{J_1(\kappa\delta)}{\kappa\delta}\nn\\
&&\quad\times\frac\partial{\partial\kappa}
\ln\kappa^2[I_0'(\kappa b)K_0'(\kappa a)-I_0'(\kappa a)K_0'(\kappa b)],\nn\\
\label{em0}
\eea
where the divergent terms arise from the large argument expansions
\begin{subequations}
\bea
I_0'(x)&\sim&\frac{e^x}{\sqrt{2\pi x}}\left(1-\frac3{8x}-\frac{15}{128x^2}+
\dots\right),\\ 
K_0'(x)&\sim&e^{-x}\sqrt{\frac{\pi}{2 x}}\left(1+\frac3{8x}-\frac{15}{128x^2}+
\dots\right).\nn\\
\eea
\end{subequations}
Inserting this into Eq.~(\ref{em0}) we obtain
\bea
\hat{\mathcal{E}}_{m=0}&\sim&-\frac1{4\pi\delta}\int_0^\infty d\kappa\,
J_1(\kappa\delta)\bigg[\kappa(b-a)+1\nn\\
&&\mbox{}+\frac38\frac1\kappa\left(\frac1b-\frac1a\right)
+\frac38\frac1{\kappa^2+\lambda^2}\left(\frac1{b^2}+\frac1{a^2}\right)
+\dots\bigg]\nn\\
 &\sim&-\frac{b-a}{4\pi\delta^3}-\frac1{4\pi\delta^2}+\frac3{32\pi\delta}
\left(\frac1a-\frac1b\right)\nn\\
&&\quad\mbox{}+\frac3{64\pi}\ln\lambda\delta\left(\frac1{a^2}+
\frac1{b^2}\right).
\eea
Here, in the last term we introduced a mass, $\kappa^2\to\kappa^2+\lambda^2$,
in order to eliminate the infrared divergence.  These terms all agree with
the corresponding terms found from the uniform asymptotic expansion by
taking $m=0$.  We might note that these terms are all independent of $\alpha$,
so cannot contribute to the torque, but for completeness we will retain them.

There is one remaining divergent term, arising from the $1/\nu^3$ term, but
here we exclude $m=0$, because that subtraction is not necessary since
the corresponding
$m=0$ contribution to the energy is already finite at $\delta=0$.  That 
curvature term is
\be
\hat{\mathcal{E}}_0\sim \frac\alpha{180\pi^2}\ln\delta\left(\frac1{b^2}
-\frac1{a^2}\right).
\ee

Let us summarize the divergent terms for the Neumann or TE modes:
\bea
\hat{\mathcal{E}}_{\rm div}&=&-\frac{A}{2\pi^2\delta^4}-\frac{P}{16\pi\delta^3}
-\frac{C}{48\pi\delta^2}\nn\\
&&\mbox{}+\frac{3}{64\pi\delta}\left(\frac1a-\frac1b\right)
-\frac{5\alpha}{1024\pi\delta}\left(\frac1a+\frac1b\right)\nn\\
&&\mbox{}+\frac{3\ln\delta/\mu}{128\pi}\left(\frac1{a^2}+\frac1{b^2}\right)
-\frac{\alpha\ln\delta/\mu}{180\pi^2}\left(\frac1{a^2}-\frac1{b^2}
\right).\nn\\
\label{ediv}
\eea
Here, we have introduced an arbitrary scale $\mu$, which will appear in
the finite part given in the next section.

\subsection{Heat-kernel expansion}
This small-$\delta$ Laurent expansion (\ref{ediv}) exactly agrees with that found by
the heat-kernel calculation of Dowker and Apps and of
Nesterenko, Pirozhenko, and Dittrich
\cite{Nesterenko:2002ng,Dowker:1995sp,apps}, who consider a wedge intercut
 with a single coaxial circular cylinder with radius $R$. 
 From the latter heat-kernel coefficients
the cylinder-kernel coefficients 
can be readily extracted \cite{Fulling:2003zx}.  The trace of the
cylinder kernel $T(t)$ is defined in terms of the eigenvalues of the
Laplacian in $d$ dimensions,
\be
T(t)=\sum_j e^{-\lambda_j t}\sim \sum_{s=0}^\infty e_s t^{s-d}+\sum_{s=d+1\atop
s-d \,{\rm odd}}f_s t^{s-d}\ln t,
\ee where the expansion holds as $t\to0$ through positive values.  The
energy is given by
\be
E(t)=-\frac12\frac\partial{\partial t}T(t),
\ee
which corresponds to the energy computed here with $\phi=0$, that is, 
time-splitting.  In view of Eq.~(\ref{relttoz}) we see that the $z$-splitting
result should be identical to that of $-\frac1{2t} T(t)$ with $t\to\delta$.
In this way we transcribe the results of Ref.~\cite{Nesterenko:2002ng} for
the outside cylinder kernel per unit length:
\bea
-\frac1{2t}T(t)&\sim&-\frac{A}{2\pi^2 t^4}-\frac{P}{16\pi t^3}
-\frac{1}{16\pi^2 t^2}
\nn\\
&&\mbox{}+\frac{3-5\alpha/16}{64\pi R t}+\frac{\ln t}{16\pi^2R^2}
\left(\frac{3\pi}8-\frac{4\alpha}{45}\right).\nn\\
\eea
This exactly agrees with Eq.~(\ref{ediv}) when $a\to R$ and $b\to \infty$ 
(except that the limits are not taken in the first two terms). 
 The reason for the factor of 2 discrepancy
in the third (corner) term is that Nesterenko et al.\ have only two corners,
not four.

\section{Extraction of finite part}\label{sec:finite}
Just as in the Dirichlet case considered in I, the divergent terms have
finite remainders, which we state here:
\bea
\hat{\mathcal{E}}_f&=&-\frac{\pi^2}{2880\alpha^3}\left(\frac1{a^2}-
\frac1{b^2}\right)-\frac{\zeta(3)}{64\pi\alpha^2}\left(\frac1{a^2}+\frac1{b^2}
\right)\nn\\
&&\mbox{}+\frac1{576\alpha}\left(\frac1{a^2}-\frac1{b^2}\right)\nn\\
&&\mbox{}+\left\{\frac3{128\pi b^2}\left[-\frac{11}{12}+\gamma+
\ln\frac{b\alpha}\mu+2\ln \mu\lambda\right]
+(b\to a)\right\}\nn\\
&&\mbox{}+\left\{\frac\alpha{\pi b^2}\left(-\frac1{180\pi}\ln
\frac{b\alpha}{\pi\mu}+\frac{1079}{69120}\right)-(b\to a)\right\}\nn\\
&&\mbox{}+\frac{29}
{46080}\frac{\alpha^2}{\pi}\left(\frac1{a^2}+\frac1{b^2}\right)\nn\\
&&-\frac{5}{12012}\frac{\alpha^3}{\pi^4}\zeta(3)\left(\frac1{a^2}
-\frac1{b^2}\right)+\hat{\mathcal{E}}_R.\label{efinite}
\eea
The last two explicitly given terms are what come from the next two terms
in the uniform expansion for $m>0$.  Note that we have made no approximation
here, we have merely added and subtracted the leading terms in the uniform
asymptotic expansion of the integrand for the energy.  The remainder, 
therefore, consists of two parts, that arising from $m=0$:
\bea
\hat{\mathcal{E}}_{R0}&=&-\frac1{8\pi}\int_0^\infty d\kappa
\,\kappa\bigg[\kappa
\frac{\partial}{\partial\kappa}\ln\kappa^2\tilde\Delta_{m=0}-\kappa(b-a)-1\nn\\
&&\quad\mbox{}-\frac3{8\kappa}\left(\frac1b-\frac1a\right)
-\frac3{8(\kappa^2+\lambda^2)}\left(\frac1{b^2}+\frac1{a^2}\right),\nn\\ 
\label{er0}
\eea
and the rest coming from the terms with $m>0$:
\bea
\hat{\mathcal{E}}'_R&=&-\frac1{8\pi b^2}\sum_{m=1}^\infty \nu^3\int_0^\infty
d\xi\,\xi^2\bigg[\hat f(\nu,\xi,a/b)\nn\\
&&\quad\mbox{}+\sum_{n=4}^{-2}\hat f_n(\nu, \xi, a/b)
\bigg].\label{erp}
\eea
Here, with the abbreviations $I=I_\nu(\nu \xi)$, $\tilde I=I_\nu(\nu \xi a/b)$,
etc., the log term is
\be
\hat f=\frac{\left(1+\frac1{\xi^2}\right)(I\tilde K'-K\tilde I')
+\frac{a}b\left(1+\frac{b^2}{a^2\xi^2}\right)(I'\tilde K-K'\tilde I)}
{I'\tilde K'-K'\tilde I'}.
\ee
The subtractions are easily read off:
\begin{subequations}
\bea
\hat f_4&=&-\frac1{\xi t}+\frac{a}b\frac1{\tilde \xi\tilde t},\\
\hat f_3&=&-\frac1{2\nu}(\xi t^2+\frac{a}b\tilde \xi\tilde t^2),\\
\hat f_2&=&\frac1{8\nu^2}\xi t^3(-3+7t^2)-\frac{a}b(\xi\to\tilde \xi),\\
\hat f_1&=&\frac1{8\nu^3}\xi t^4(-3+20t^2-21t^4)+\frac{a}b(\xi\to\tilde \xi
),\\
\hat f_0&=&\frac1{5760\nu^4}\xi t^5(-2835+39105 t^2-99225t^4\nn\\
&&\mbox{}+65835 t^6)-\frac{a}b(\xi\to\tilde \xi),\\
\hat f_{-1}&=& \frac1{128\nu^5}\xi t^6(-108+2616 t^2-11728 t^4
\nn\\&&\quad\mbox{}+17640 t^6
-8484 t^8)+\frac{a}b(\xi\to\tilde \xi),\\
\hat f_{-2}&=&\frac1{32560\nu^6}\xi t^7(-598185+22680945 t^2\nn\\
&&\quad\mbox{}-156073050t^4+393353730 t^6
\nn\\&&\quad\mbox{}-415212525 t^8+156010365 t^{10})
-\frac{a}b(\xi\to\tilde \xi).\nn\\
\eea
\end{subequations}
The last two subtractions, and the associated terms in Eq.~(\ref{efinite}),
 are not necessary, but they improve convergence.

\section{Numerics}\label{sec:num}
The extraction of the finite part follows the same procedure described  in I.
The total finite energy given in Eq.~(\ref{efinite}) 
is the sum of the explicitly given  finite terms plus the remainder:
\be
\hat{\mathcal{E}}_f=\sum_{n=4}^{-2}\hat{\mathcal{E}}^f_n+\hat{\mathcal{E}}_R,
\label{ef} 
\ee
where 
$\hat{\mathcal{E}}_R$ is the sum of Eqs.~(\ref{er0}) and (\ref{erp}).

\begin{figure}
\includegraphics[scale=.65,clip]{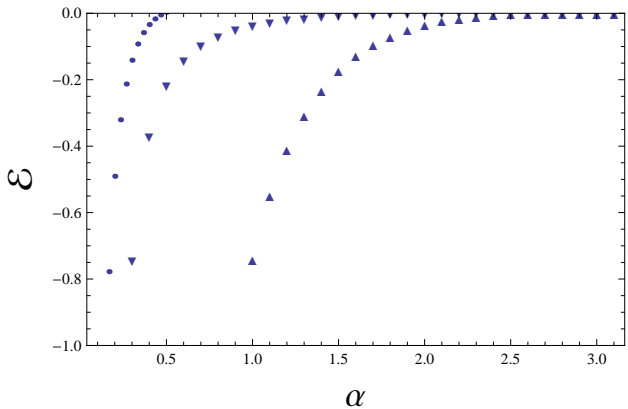}
\caption{\label{figte} Renormalized energy for TE modes for $a/b=0.1$
(dots), $a/b=0.5$ (inverted triangles), and $a/b=0.9$ (triangles). Here
the energies per unit length, in units of $1/b^2$, are plotted as a function
of the wedge angle $\alpha$.}
\end{figure}

The total energy becomes a linear function of $\alpha$ for sufficiently
large wedge angles. But because of the logarithmically  divergent parts 
in the energy, such linear terms are undetermined.
That is, we can add to the energy an arbitrary counter term of the form
\be
\hat{\mathcal{E}}_{ct}=A+B\alpha.
\ee
We subtract off the linear behavior found numerically 
from Eq.~(\ref{ef}), because
the energy should approach zero for sufficiently (but not very) large
$\alpha$.  In this way, we get the Neumann (TE)  energies  
seen in Fig.~\ref{figte}, very similar to what we found for the Dirichlet
(TM) contribution.

\begin{figure}
\includegraphics[scale=.65,clip]{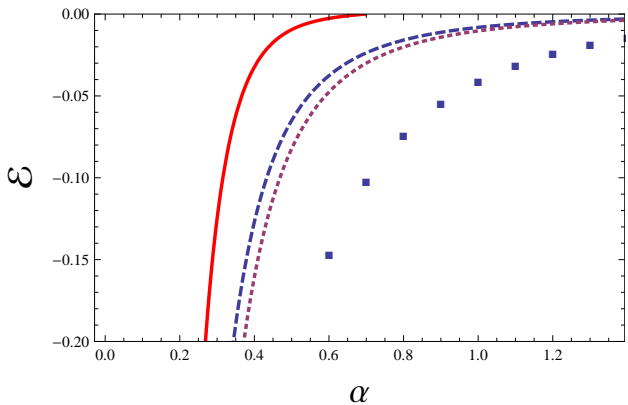}
\caption{\label{tetmc} Comparison of the renormalized energy per length for TE 
modes (squares) and TM modes (upper red curve), 
with the Casimir limit (\ref{caslim}) (blue dashes)
and the PFA approximation [first term in Eq.~(\ref{efinite})] (red dots),
 the PFA being larger than the Casimir limit in magnitude by 27\%,
for $a/b=0.5$. The TE mode is always much larger in magnitude than the TM 
contribution.}
\end{figure}

The TE and TM contributions are both displayed in Fig.~\ref{tetmc},
as well as the result expected for either TE or TM modes
for parallel plates, which is approached as $\alpha\to0$:
\be
\mathcal{E}_C=-\frac{\pi^2}{180 b^2}\frac{1-a/b}{(1+a/b)^3}\frac1{\alpha^3}.
\label{caslim}
\ee
This formula is valid in the regime $\alpha\ll1$, $1-a/b\ll1$, in which case
it agrees with the leading finite term in Eq.~(\ref{efinite}), which is
the same as the $\alpha^{-3}$ term for the Dirichlet case. 
Those leading terms are, in fact, the proximity force approximation (PFA). 
Figure \ref{tetmc} shows that this limit is indeed approached for both TE and
TM modes, from opposite sides,
but that the TE mode is always considerably larger in magnitude than the TM
mode, which is a phenomenon observed previously in a related context
\cite{Milton:2004ya}.
\section{Conclusions}\label{sec:concl}

Because of curvature divergences, it is impossible to extract a unique
finite part of the energy.  However, the divergences are all constant or
linear in the wedge angle $\alpha$.  Therefore, we can renormalize the
energy by subtracting the linear dependence for large angles, to
impose a physical requirement that the
energy go to zero when the separation between the wedge planes is large.
The resulting energy is completely finite,
independent of regularization scheme, and exhibits no torque anomaly:
\be
\tau(\alpha)=-\frac\partial{\partial\alpha}\mathcal{E}(\alpha).
\ee
These results, of course, are consistent with, and generalize to
electromagnetism, the annular piston work 
of Ref.~\cite{Milton:2009bz}.   It is remarkable how similar
the electromagnetic calculation is to that for the Dirichlet scalar.

So, as with the scalar, Dirichlet, case, there is no sign of a torque
anomaly.  Here, this is even more surprising, because in the Dirichlet
situation, the anomaly is manifested by linear terms in $\alpha$ in the
energy, which would be canceled by the corresponding exterior ($\theta
\in [\alpha,2\pi]$) contribution for an annular piston, as well as being
removed by our ``renormalization'' procedure. 
As emphasized in Ref.~\cite{fulling12},  for the 
electromagnetic wedge, there is an additional anomalous term in the energy
 $\sim \alpha^{-1}$ \cite{deutsch},
which would not disappear if the exterior contribution were included, and
should not be removed by renormalization.
The reason we do not see this effect here will be explored further
as we study the local regulated stress tensor. 

To summarize, in these two papers, we have explored the torque $\tau$ (per unit length)
on one side of an annular
sector, formed by the intersection of two planes, and two coaxial cylinders.
The question we asked was whether the torque was somehow anomalous, in that 
\be
\tau\ne-\frac\partial{\partial \alpha}\mathcal{E}?
\ee
Here $\mathcal{E}$ is the energy (per unit length) contained within the sector,
and $\alpha$ is the dihedral angle between the planes. In the first paper I, the quantum
vacuum energy and torque were computed for a massless scalar field subject to 
Dirichlet boundary conditions on all the surfaces, and in the present paper, the
boundaries are perfect conductors, and the fluctuating field is the electromagnetic
one.  In both cases we computed the divergent and finite parts of the energy,
obtained by point-splitting in either the (Euclidean) time or the axial direction.
The physical normalization requirement that the energy of the annular sector
go to zero for sufficiently large wedge angles, allows us to define a finite,
non-anomalous renormalized energy.  The possibility of doing so, however, 
depends on the existence of an inner cylindrical boundary.  Without that boundary
it is not possible to define a torque or an energy, and ambiguities such as
the torque anomaly can appear.
\acknowledgments
We thank the U.S. National Science Foundation and the Julian Schwinger
Foundation for partial support of this work.  We thank our many collaborators,
especially Jeffrey Bouas, Iver Brevik, Stuart Dowker, Stephen Fulling, 
Stephen Holleman, K. V. Shajesh, and Jef Wagner,  for helpful discussions.
F. K. thanks the Homer L. Dodge Department of Physics and Astronomy of the
University of Oklahoma for its hospitality during the period of this work.

\end{document}